\documentclass[twocolumn,prl,floatfix,superscriptaddress,amsmath,amssymb,A4paper]{revtex4-2}
\usepackage{graphicx, ulem}
\usepackage{amssymb}
\usepackage{bm}
\usepackage{textcomp}
\usepackage{xfrac}
\usepackage{yfonts}
\usepackage{color}
\usepackage{float}
\usepackage{braket}

\graphicspath{{pics}}
\begin{document}

\title{Terahertz resonant emission by optically excited infrared-active shear phonons in KY(MoO$_4$)$_2$}

\author {D. Kamenskyi}
\affiliation {Institute of Optical Sensor Systems, German Aerospace Center (DLR), Rutherfordtstr. 2, 12489 Berlin, Germany}
\affiliation {Department of Physics, Humboldt-Universität zu Berlin, Newtonstr. 15, 12489 Berlin, Germany}
\affiliation {Experimental Physics V, Center for Electronic Correlations and Magnetism, Institute of Physics, University of Augsburg, 86159 Augsburg, Germany}
\author {K. Vasin}
\author {L. Prodan}
\affiliation {Experimental Physics V, Center for Electronic Correlations and Magnetism, Institute of Physics, University of Augsburg, 86159 Augsburg, Germany}
\author{K. Kutko}
\author{V. Khrustalyov}
\affiliation{B. Verkin Institute for Low Temperature Physics and Engineering of the National Academy of Sciences of Ukraine, Nauky Avenue 47, 61103 Kharkiv, Ukraine}
\author {S. G. Pavlov}
\affiliation {Institute of Optical Sensor Systems, German Aerospace Center (DLR), Rutherfordtstr. 2, 12489 Berlin, Germany}
\author{H.-W. H\"ubers}
\affiliation {Institute of Optical Sensor Systems, German Aerospace Center (DLR), Rutherfordtstr. 2, 12489 Berlin, Germany}
\affiliation {Department of Physics, Humboldt-Universität zu Berlin, Newtonstr. 15, 12489 Berlin, Germany}

\begin{abstract}

Generation of the monochromatic electromagnetic radiation in the terahertz (THz) range of frequencies for many decades remanes a chellenging task. Here we demonstrate the emission of monochromatic sub-THz radiation by coherent optical phonons in dielectric material KY(MoO$_4$)$_2$. The layered crystal structure of KY(MoO$_4$)$_2$ leads to infrared-active shear lattice vibrations with energies below 3.7 meV, which corresponds to the frequencies low than 900 GHz where solid-state compounds based monochromatic radiation sources are rare. Coherent infrared-active optical phonons are excited by broadband THz pulses lasting for tens of picoseconds and re-emit narrow-band sub-THz radiation pulses with a decay time of 33 picoseconds, which is exceptionally long for the oscillators with frequencies below 1 THz. Such a long coherent emission allows for the detection of more than 50 periods of radiation with frequencies of 568 and 860 GHz. The remarkably long decay time together with the chemical stability of the employed materials suggest a variety of possible applications in THz laser technology.

 \end{abstract}

\maketitle

Last decades significant achievement in the Time-Domain Terahertz Spectroscopy (THz-TDS) boost interest to THz research and its application \cite{Mittelman}. THz-TDS method is based on electromagnetic transients opto-electronically generated by ultrashort, usually femtosecond, laser pulses. The single-cycle bursts of electromagnetic radiation typically have duration about 1~ps with the spectral density enhanced between 100 GHz and 5~THz~\cite{Nuss}. It is well known that a time-varying electric dipole moment emits electromagnetic wave on the oscillation frequency of the dipole~\cite{Nuss}. For the first time, the photo-Dember effect of electromagnetic wave re-emission by coherent phonons in the THz range of frequencies has been reported in tellurium single-crystal~\cite{emission1995}. Recently, it was also reported in hybrid perovskites \cite{emission2018}, charge-density wave system K$_{0.3}$MoO$_3$ ~\cite{emission2019}, and  topological material TaAs where it offers opportunities for THz emission with polarization control~\cite{emission2020}. The recent overview of the THz re-emission by optically pumped solids is given elsewhere~\cite{review2022}. Generally, the photo-Dember THz re-emission is broadband with the typical spectrum width above 100~GHz.

Despite the achieved progress, the generation of the monochromatic THz radiation, remains a challenging task. However for such applications as a THz imaging \cite{imaging2014}, THz-driven particle acceleration~\cite{acceleration2015} or ultrafast phase-changes~\cite{phase_ch2012}, narrow-band THz pulses is crucial. Lithium niobate \cite{LN} as well as other materials (semiconductors \cite{semiconductors} and organic crystals \cite{organics2015}) have been reported as perspective materials for generating high-energy narrow-band THz pulses with the typical bandwidth of about 1\%. Here we report observation of a narrow-band (full width on half maximum~$\Delta\nu < 4$ GHz or 0.5\%) re-emission of THz electromagnetic pulses by KY(MoO$_4$)$_2$ single crystals pumped by a broadband THz excitation pulse in a conventional THz-TDS setup. KY(MoO$_4$)$_2$ is a solid dielectric material whose dipole-active shear lattice vibrations lie in THz frequencies range. The observed length of the re-emitted pulses of quasi-monochromatic electromagnetic radiation is over a hundred of picoseconds (tens of optical cycles), that is longer than the photon transient time in the used crystals. We explain this emission as resonant re-emitted light due to photo-excitation of infrared-active vibrational shear modes in KY(MoO$_4$)$_2$.

\begin{center}
\begin{figure*}
\centering
\includegraphics[width=0.9\textwidth, keepaspectratio]{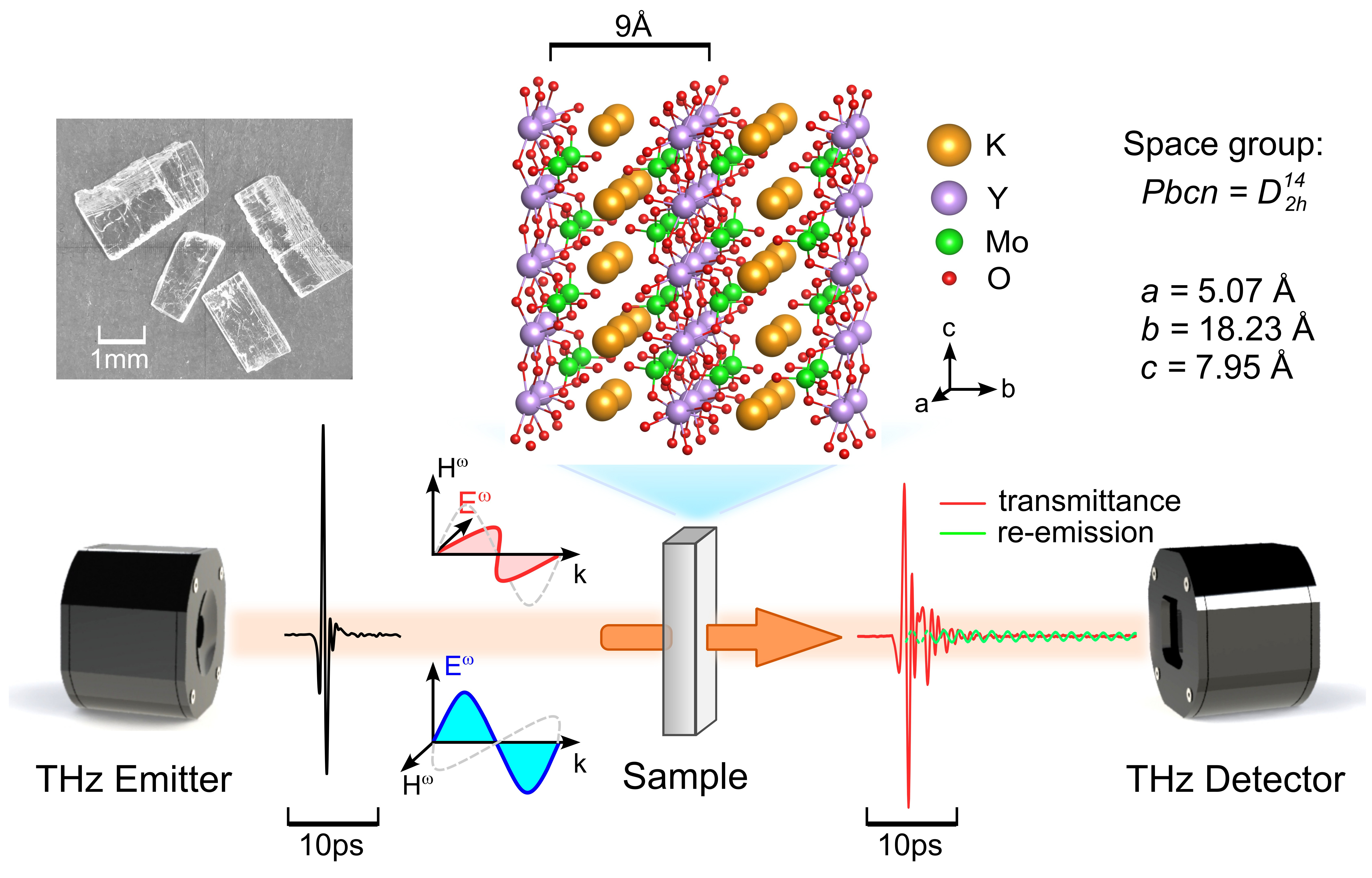}
\caption{Top. A photograph and the crystallographic structure of KY(MoO$_4$)$_2$ single crystal.\\ 
Bottom. A schematic representation of THz-TDS setup: black trace shows THz waveform of incident light (from THz emitter). Red and green curves show the components of the beam which enters the detector. The cryogenic part of the setup is omitted for clarity. $E^\omega$ stands for the electric field magnitude of electromagnetic wave.}
\label{structure}
\end{figure*}
\end{center}

KY(MoO$_4$)$_2$ belongs to a series of molybdate compounds with general chemical formula \textit{MR}(MoO$_4$)$_2$, where \textit{M}$^+$ is an alkali metal ion and \textit{R}$^{3+}$ is a rare-earth or Y$^{3+}$ ion. These are optically transparent dielectric compounds with a high permittivity ($\epsilon > 15$) and orthorhombic structure \textit{Pbcn} (\textit{D}$_{2h}^{14}$) which is formed by [\textit{R}(MoO$_4$)$_2$]$^-$ layers in \textit{ac} plane coupled via $M^+$ ions along the \textit{b} direction (see crystallographic structure in Figure~\ref{structure})~\cite{phonons}. The detailed description of the synthesis is given in Ref. \cite{Structure}. KY(MoO$_4$)$_2$ single crystals used for this study have been grown in the Institute for Low Temperature Physics (Kharkiv, Ukraine). The compositional analysis of the crystals was done using scanning electron microscopy in combination with the energy-dispersive X-ray spectroscopy and elemental mapping.

In recent years rare-earth based molybdates attract attention due to a giant rotational magnetocaloric \cite{KEr_caloric} and massive magnetostriction \cite{Striction} effects induced by large anisotropy of the \textit{R}$^{3+}$ ion. Here we exploit a large anisotropy of the crystal lattice of KY(MoO$_4$)$_2$ for the emission of THz radiation. In our experiments we used the plate-shaped single crystals (Figure~\ref{structure} left) with the crystallographic \textit{b} axis perpendicular to the plates. Due to a weak bonding between [\textit{R}(MoO$_4$)$_2$]$^-$ layers, single crystals cleave along $ac$ plane.

THz-TDS spectra were measured using a table-top time-domain terahertz platform TeraFlash (TOPTICA Photonics AG). In these experiments, the radiation was propagating along the \textit{b}-axis of the crystal, and linearly polarised along the \textit{a} and \textit{c}.

Figure~\ref{structure} shows a sketch of the experimental setup, image of samples and the crystallographic structure of  KY(MoO$_4$)$_2$ crystal. The specimen was mounted to a brass holder with 4~mm aperture inside the Spectromag cryostat (Oxford Instruments) (which is omitted in the Figure~\ref{structure} for clarity) with insert which allows temperature variation between 3 and 300 K. The black curve shown before the sample in Figure~\ref{structure} represents a temporal profile of electric field during the THz pulse measured when the radiation passed the empty aperture (excitation pulse), and curves after the sample show the profile measured when the pulse passed the sample. The pulse after the sample contains contributions of the distorted initial TDS pulse (red curve) and the radiation re-emitted by the crystal (green curve), further we discuss the microscopic mechanism of this re-emission. The square of the ratio of the Fourier transforms of the measured transients returns the sample's transmittance spectrum (transmitted energy). Taking this ratio as a transmittance, we assume that the sample does not affect significantly the efficiency of the light collection by the detector.

\begin{center}
\begin{figure}
\centering
\includegraphics[width=0.5\textwidth, keepaspectratio]{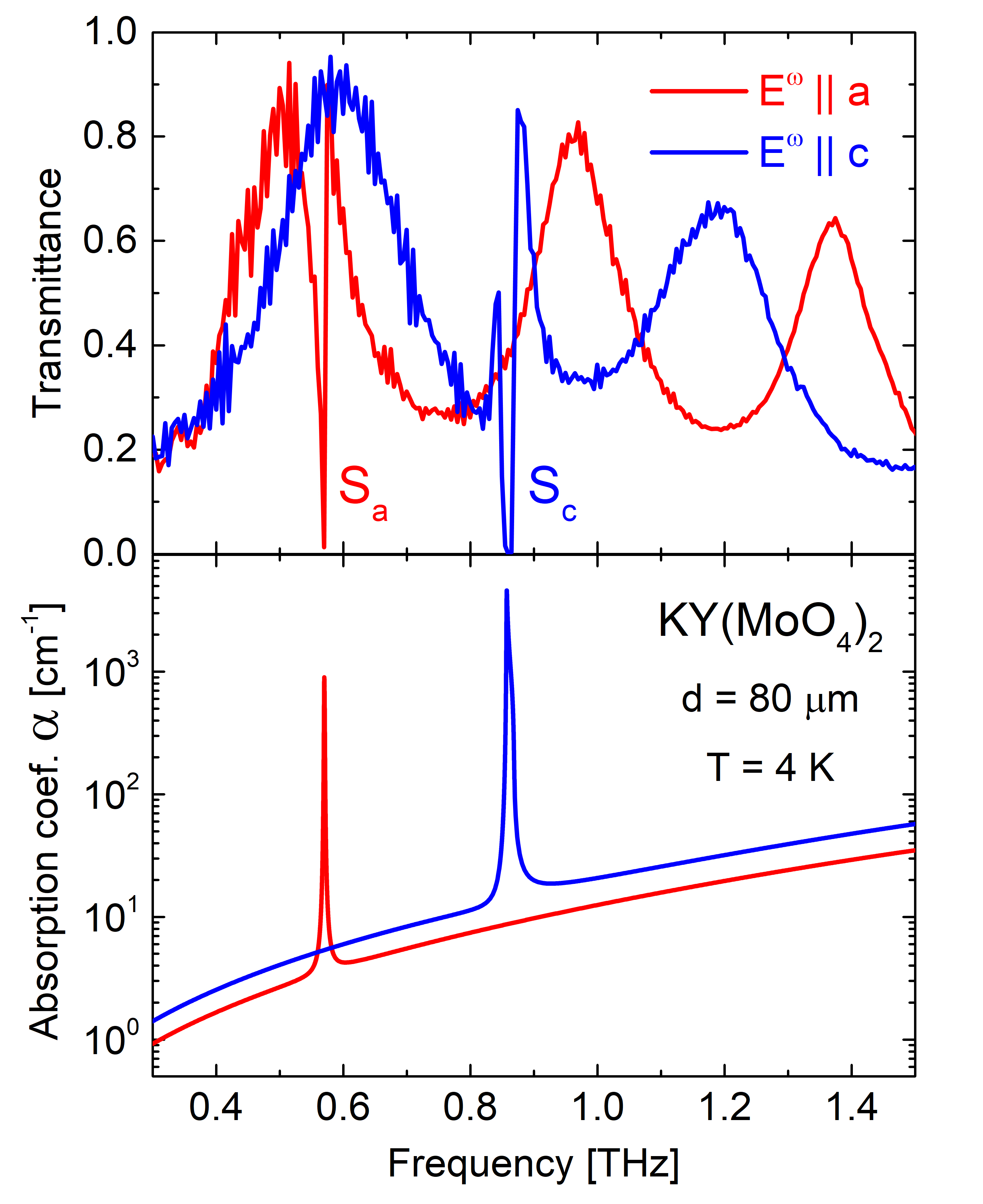}
\caption{Top. Transmittance spectra of 80 µm thick KY(MoO$_4$)$_2$ sample for $E^\omega \parallel a$  (red) and $E^\omega \parallel c$ (blue) polarisations. Multiple reflections within the plane-parallel sample result in Fabry–Perot type modulation of spectra. \\Bottom. Frequency dependence of the absorption coefficient $\alpha(\omega)$ obtained using REFFIT script \cite{REFFIT}.}
\label{spectra}
\end{figure}
\end{center}

Figure~\ref{spectra} top displays the transmittance spectra of 80~$\mu$m thick KY(MoO$_4$)$_2$ single crystal at frequencies between 0.3 and 1.5~THz measured at 4~K. Red and blue curves correspond to the THz wave polarised along $a$ and $c$ crystallographic axes respectively ($E^\omega \parallel a$ and $E^\omega \parallel c$, where $E^\omega$ is the electric field magnitude of the electromagnetic wave). Sharp peaks $S_a$ and $S_c$ at 568 and 860 GHz (phonon energy equivalents are 2.35 and 3.56~meV) respectively, are due to dipole active shear lattice vibrations, when [Y(MoO$_4$)$_2$]$^-$ and K$^+$ layers move as a whole in $ac$ plane as has been shown earlier \cite{phonons}. A large mass of [Y(MoO$_4$)$_2$]$^-$ layers together with weak bonding between them causes the frequency of these vibrations are below 1 THz, while typical optical phonons in solids have frequencies above 3 THz~\cite{Yoshida}. Remarkably, the  $S_c$ absorption is about 10 times stronger then $S_a$. This make the sample opaque for $E^\omega \parallel c$ radiation with frequencies in a vicinity of $S_c$ peak (Figure~\ref{spectra}), and cause a peculiarity in the re-emission spectrum of $S_c$ phonon which we are going to discuss further in the text.

Multiple reflections of the radiation within the plane-parallel sample result in interference fringes of the transmission spectra (Figure~\ref{spectra} top). The periodicity of the fringes determined by the thickness of the crystal, $d$, and the refractive index, $n$, of the sample. We used REFFIT script \cite{REFFIT} to determine the complex dielectric function of the material, $\hat{\epsilon}(\omega) = \epsilon_1-i\epsilon_2$, and, consequently, frequency dependence of the refractive index, $n$, the optical extinction coefficient, $\kappa$, and the absorption coefficient $\alpha(\omega) = 2\kappa\omega/c_0 = 2\omega/c_0\sqrt{1/2\cdot(\epsilon_1^2+\epsilon_2^2)^{1/2}-\epsilon_1/2}$ (Figure~\ref{spectra} bottom) where $c_0$ is the speed of light in vacuum. According to the Beer–Lambert law,  $I(d)=I_0 e^{-\alpha\cdot d}$ ($d$~is the sample thickness), $1/\alpha$ corresponds to the sample thickness when the light intensity, $I$, decays in $e$ times.

Extremely low width of the peaks (for $E^\omega \parallel a$ it is below reliable resolution limit of 5~GHz) reflects exceptionally large lifetime of such a dipole. To the best of our knowledge, these are the sharpest phonon peaks ever observed below 1 THz. The lifetime could be estimated as $\tau=(\pi\Delta\nu)^{-1} \approx 100$ ps \cite{width1,width2} (where $\Delta\nu$ is the full width at half maximum of $S_a$ mode in the $\alpha(\omega)$ spectrum, Figure \ref{spectra} bottom).  Thus, $\tau$ of $S_a$ phonon is an order of magnitude exceeds usual phonons lifetime (typically $1-10$~ps) and the duration of the excitation THz pulse in our setup ($\sim 5$~ps). For the $S_c$ phonon such estimation is not possible because of the strong absorption in a vicinity of $S_c$ and the transmitted signal between 857 and 863 GHz is below the noise level of the detector and the peak is distorted (later in the text we call such distortion of the peak $saturation$).

\begin{center}
\begin{figure*}
\centering
\includegraphics[scale=0.6, keepaspectratio]{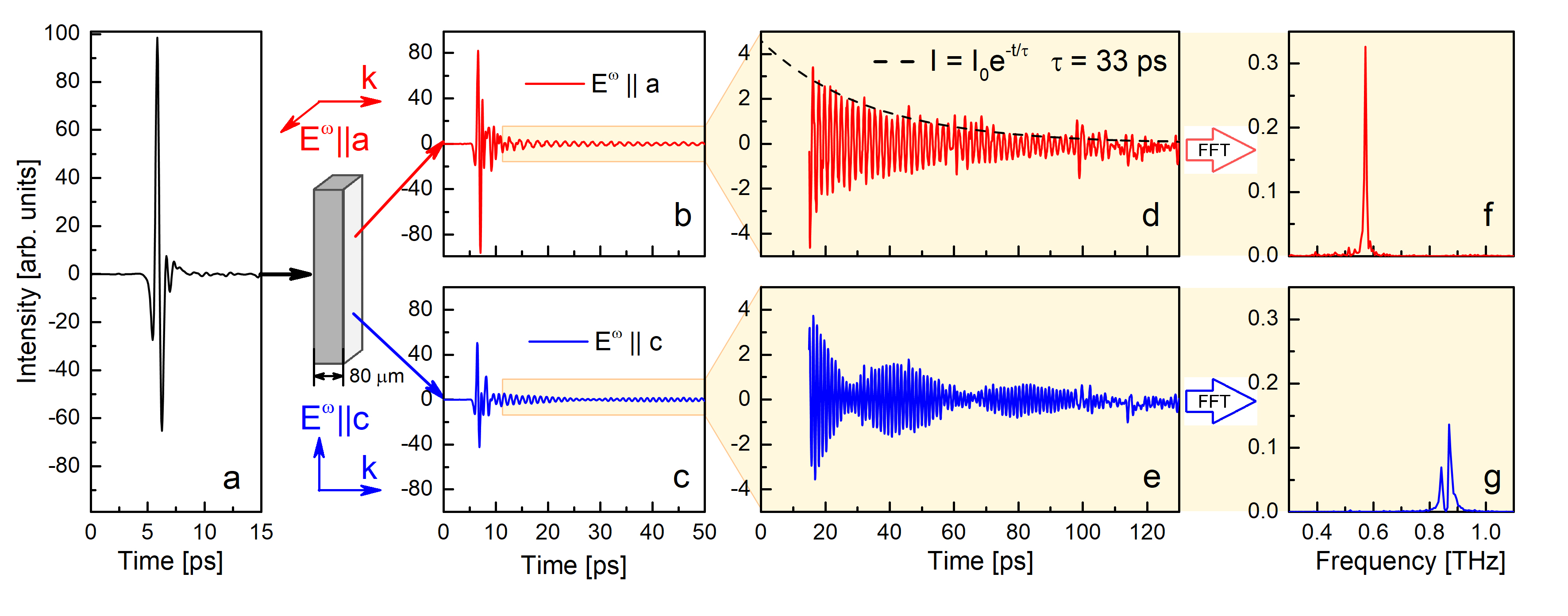}
\caption{(a) The electric field waveform of the THz pulse before the sample.\\ (b, c) Waveforms of the THz pulse passed the sample for polarisations along $a$ (red) and $c$ (blue) axes.\\ (d, e) Zoomed waveforms for the $a$ (red) and $c$ (blue) polarisations 10 ps after the start of the pulse. \\(f, g). Squared FFT of the waveforms shown in \ref{time}d and \ref{time}e, respectively.}
\label{time}
\end{figure*}
\end{center}

Figure~\ref{time}a shows the electric field waveform of the broadband THz pulse which reaches the detector via an empty aperture. This signal decays within 4-5~ps and, obviously, corresponds to the pulse before it passed the sample (losses in optical elements are assumed to be the same for the beams passing the cryostat with and without sample). The waveforms which passed 80 µm thick sample (Figures 3b, 3c for $E^\omega \parallel a$ and $E^\omega \parallel c$ respectively) exhibit scintillating signals on phonon frequencies for tens of picoseconds. Note, the transmitted signal is also delayed by 1 ps due to a large refractive index of KY(MoO$_4$)$_2$, $nd/c_0 \sim$~1~ps ($n \approx 4$ below 1 THz). For $E^\omega \parallel a$, the waveform shows oscillations with $S_a$ phonon frequency with the decay an order of magnitude longer then the initial pulse (highlighted by the yellow background in \ref{time}b and zoomed in Figure \ref{time}d). This long extended emission tail is a manifestation of the electromagnetic wave re-emission by coherent phonons in KY(MoO$_4$)$_2$. Dashed line in Figure \ref{time}d shows the exponential decay of radiation intensity fitted by $\exp(-t/\tau_e)$ where $\tau_e$ = 33~ps is the decay time. Such a long decay and significant intensity of the signal allow the observation of tens of optical periods of monochromatic electromagnetic wave. Figure \ref{time}f shows the Fourier transformation (FFT) of the signal from Figure \ref{time}d. As one can see, 10 ps after the excitation pulse, the electromagnetic radiation with $E^\omega \parallel a$ behind the KY(MoO$_4$)$_2$ sample is practically monochromatic light with the frequency exactly the same to the absorption frequency of $S_a$ phonon. In contrast to $E^\omega \parallel a$, the waveform for $E^\omega \parallel c$ (Figure \ref{time}c) after main peak exhibits a beating of two frequencies (Figure~\ref{time}e) around the resonance frequency of $S_c$ phonon (see FFT in Figure~\ref{time}g). Later we show that this beating is a manifestation of the saturation observed in the transmittance spectra (Figure \ref{spectra} top, blue curve). The strong intensity of $S_c$ phonon prohibits the propagation of resonance frequency in material while the lattice vibrations on frequency near by the resonance efficiently re-emit the electromagnetic radiation.

\begin{center}
\begin{figure*}
\centering
\includegraphics[scale=0.3, keepaspectratio]{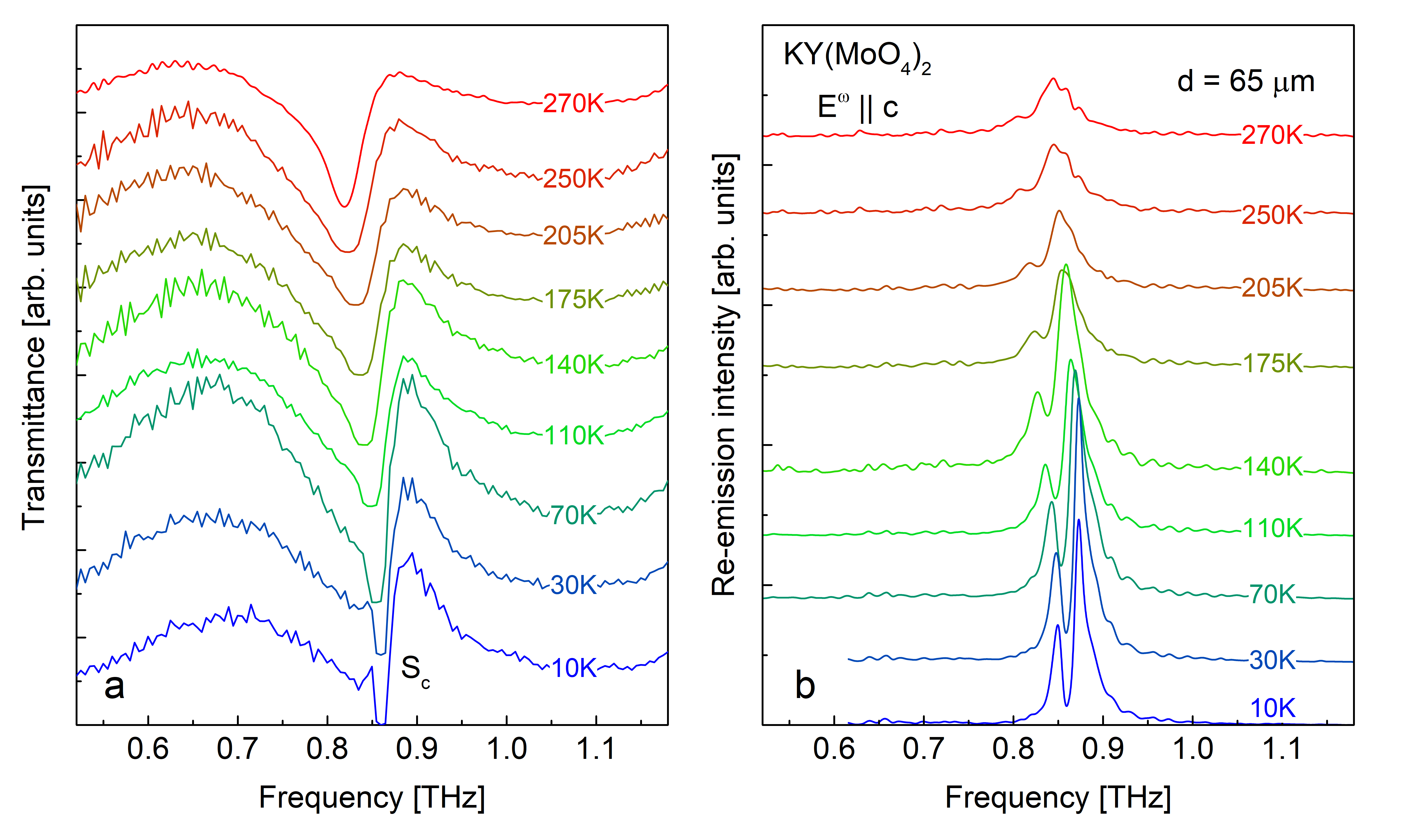}
\caption{Temperature evolution of the transmittence (a) and re-emission (b) spectra for the 65$\mu m$ thick KY(MoO$_4$)$_2$ sample at $E^\omega \parallel c$. Here, we call the FTT as “re-emission” after 10 ps delay from the pump pulse front.}
\label{temperature}
\end{figure*}
\end{center}

In order to confirm that splitting of the re-emission peak for $E^\omega \parallel c$  is caused by the strong intensity of $S_c$ phonon, we study the temperature transformation of the spectra. Figure~\ref{temperature} displays the evolution of transmittance (left) and re-emission (right) spectra of KY(MoO$_4$)$_2$ for $E^\omega \parallel c$ upon a cooling. The re-emission spectra were obtained in the same manner as in Figures~\ref{time}f and \ref{time}g. As temperature decreases the phonon peak $S_c$ gets narrow and grows in amplitude which leads to the development of the re-emission peak splitting simultaneously with saturation of the absorption peaks.  As one can see Figure~\ref{temperature}, the re-emission spectrum at room temperature (270~K) have monochromatic lorentzian shape. Upon a cooling the frequencies where the absorption is strongest are missing in re-emission spectrum which leads to a spectral dip in the emission peak. We note, that remarkably strong re-emission at room temperature significantly enlarges the application perspectives.

\begin{center}
\begin{figure}
\centering
\includegraphics[scale=0.3, keepaspectratio]{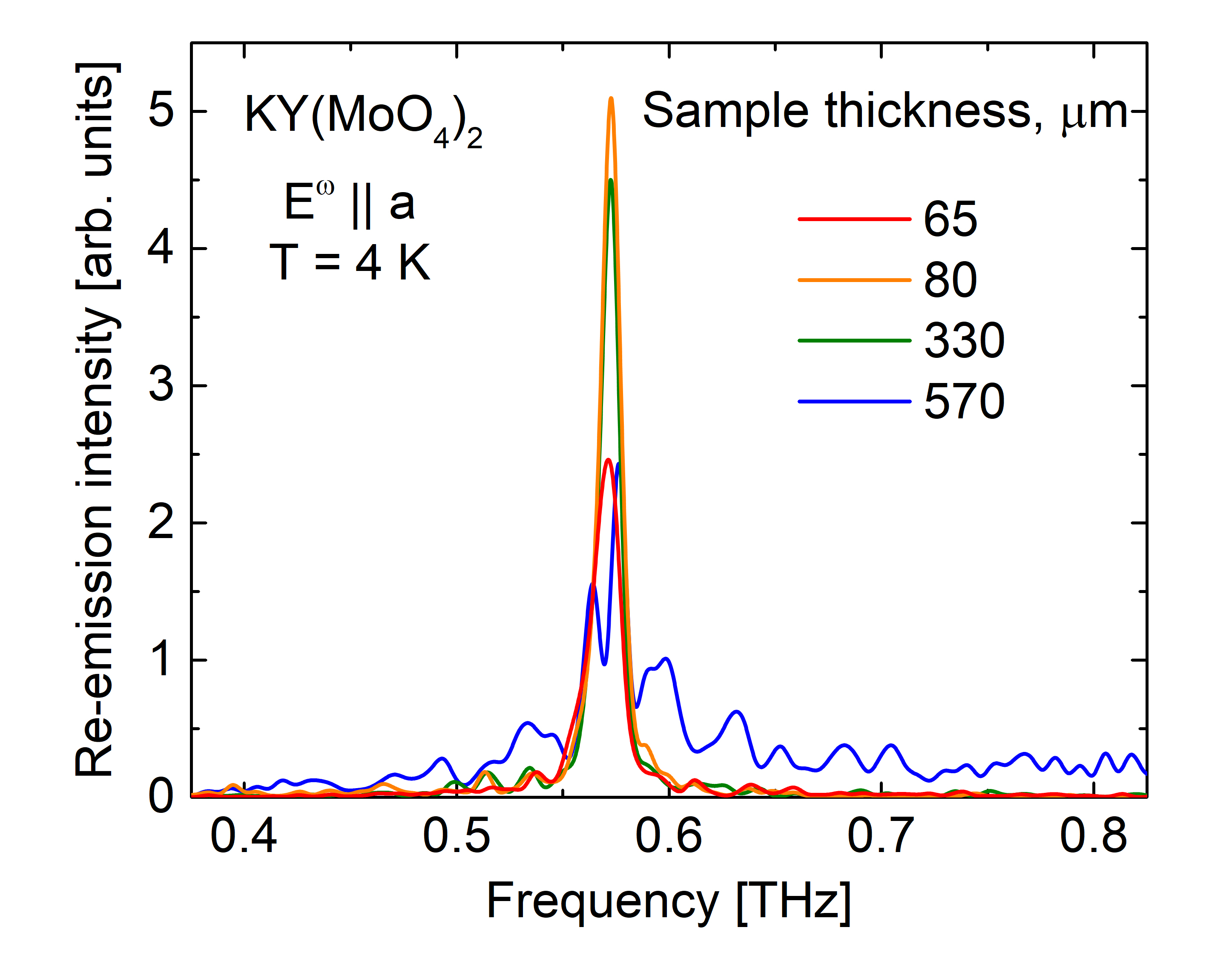}
\caption{The thickness evolution of the re-emission spectra of KY(MoO$_4$)$_2$ at 4 K and $E^\omega \parallel a$ .}
\label{thickness}
\end{figure}
\end{center}

Similar peak splitting, we observe in a thickness dependence of the re-emission spectrum (Figure~\ref{thickness} ($E^\omega \parallel a$)). One can see that maximal re-emission intensity of the 65$\mu$m thick sample is significantly smaller than for 80$\mu$m one. These thicknesses is not far from the value of $1/\alpha_a \approx 10~\mu$m at $S_a$ frequency. When the sample thickness grows further, the intensity of the re-emission decays, because obviously the radiation spend more time inside the material and the energy dissipate into other optical and acoustic phonons. When $d \gg 1/\alpha$ by an order of magnitude the re-emission intensity on the central frequency decays, while on frequencies near by the resonance the re-emission remains efficient as it has been described above for the $E^\omega \parallel c$ polarisation. This leads to the beating in the time-domain  spectra (Figure~\ref{time}e) and to the spectral dip in the re-emission peak (Figure~\ref{time}g).


Beating shapes of TDS transients propagating through the media with strong, spectrally narrow absorption peaks have attracted attention of many theoretical groups in past. Different phenomena, such as so-called Sommerfield-Brillouin forerunners \cite{forerunners-old}, dynamical beats, and optical precursors generated by the incident step-modulated pulses, were utilized to explain similar experimental observations. We point a comprehensive work by Bruno and Bernard \cite{forerunners} focused on the response function of the dielectric medium with a single absorption line. It outlines the conditions for observing dynamical beats: $1 \ll \alpha d \ll \nu_0 / \Delta\nu$, where $\nu_0$ represents the central frequency of the absorption line, and $d$ stands for the thickness of the slab. In our case, for an 80~µm sample at 4~K, this results are $1 \ll 7 \ll 475$ for the $S_a$ phonon and $1 \ll 37 \ll 329$ for the $S_c$ phonon. Thus, for both polarisation the condition is met. However, detailed discussion of the microscopic mechanism of the beating formation would require significantly higher spectral resolution, which is not feasible for the current TDS spectroscopy and goes beyond the scope of this study, where we focus on the optimal conditions for the monochromatic THz re-emission.

Thus, in KY(MoO$_4$)$_2$ the radiation energy absorbed by the $S_a$ phonon during the initial THz excitation pulse ($\sim$~5~ps) is not converted to the sample heating via phonon-phonon interactions (which is a common relaxation channel for the lattice excitations in solids) due to limited coupling of the share vibrations to other phonons. Instead, the most of these energy is re-emitted via mechanism of classical time-varied dipole by the coherent dipole active lattice vibrations as the monochromatic THz radiation \cite{Nuss}. Such coherent phonon re-emission has been reported previously in a different classes of materials ranging from elemental semiconductor tellurium \cite{emission1995} to hybrid perovskites \cite{emission2018} and topological material TaAs \cite{emission2020} but usually on frequencies above 1 THz and in rather broad spectrum. In our experiments, however, exceptionally low energy and long phonons lifetime allows to detect the extremely narrow-band re-emissions with central frequencies of 568~GHz and 860~GHz for $E^\omega \parallel a$ and $c$ crystallographic axes respectively with the decay time of 33~ps. For $E^\omega \parallel a$ we found that the re-emission last for more than 100~ps or more then 50 optical cycles. These unique characteristic together with chemical durability make KY(MoO$_4$)$_2$ very attractive for variety of THz applications, ranging from a conventional magnetic resonance spectroscopy \cite{FLARE} to narrowband THz pulses generation for novel THz-driven electron acceleration systems \cite{applications}.

Other materials from the family of \textit{MR}(MoO$_4$)$_2$ exhibit rather similar phonon spectra below 1~THz. All these compounds have $S_a$ and $S_c$ phonons similar to discussed above with small frequency variation due to different masses of \textit{R}$^{3+}$ and \textit{M}$^+$ ions~\cite{phonons}. Particularly, similar time traces we observed in KTm(MoO$_4$)$_2$ and KEr(MoO$_4$)$_2$ compounds. This create a versatile ground for the development of sub-THz radiation sources. An additional significant advantage is the chemical stability of these materials: they may be kept at the ambient laboratory conditions for decades without changing composition, crystallographic structure and other physical properties. Our experimental results have revealed no difference between samples and obtained in 2024 and in 1980th. \textit{MR}(MoO$_4$)$_2$ compounds are resistant to standard  solvents such as water, acetone, alcohol, which significantly simplify a working protocol. 
 
Acknowledgment 

The authors thank N. Stojanovic for useful discussions and suggestions. DK funded by the Deutsche Forschungsgemeinschaft (DFG, German Research Foundation) -- 497756533, KV and LP acknowledge funding from the Deutsche Forschungsgemeinschaft (DFG, German Research Foundation) TRR 360-492547816.

\end{document}